\documentclass[preprint,12pt]{elsarticle}
\usepackage{graphicx}
\usepackage{subeqn}
\usepackage{amssymb}
\journal{Physica A}

\usepackage{subeqn}

\usepackage[usenames]{color}
\usepackage{ulem} 

\newcommand{\bk}{{\bf k}}

\newcommand{\bx}{{\bf x}}

\newcommand{\cE}{\mathcal{E}}

\newcommand{\cL}{\mathcal{L}}
\newcommand{\cZ}{\mathcal{Z}}

\begin{document}

\author[dva]{D.V. Anghel}
\ead{dragos@theory.nipne.ro}

\author[bltp,iapm]{A.S. Parvan}
\ead{parvan@theor.jinr.ru}

\author[bltp,iapm]{A.S. Khvorostukhin}
\ead{hvorost@theor.jinr.ru}


\address[dva]{Department of Theoretical Physics, Horia Hulubei National Institute of Physics and Nuclear Engineering, 407 Atomistilor street, P.O.BOX MG-6, RO-077125, M\u agurele, Jud. Ilfov, Romania}

\address[bltp]{Bogoliubov Laboratory of Theoretical Physics, Joint Institute for Nuclear Research, 141980 Dubna, Russian Federation}

\address[iapm]{Institute of Applied Physics, Moldova Academy of Sciences, MD-2028 Chisinau, Republic of Moldova}


\begin{abstract}
The effect of statistics of the quasiparticles in the nuclear matter at extreme conditions of density and temperature is evaluated in the relativistic mean-field model generalized to the framework of the fractional exclusion statistics (FES). In the model, the nucleons are described as quasiparticles obeying FES and the model parameters were chosen to reproduce the ground state properties of the isospin-symmetric nuclear matter. In this case, the statistics of the quasiparticles is related to the strengths of the nucleon-nucleon interaction mediated by the neutral scalar and vector meson fields. The relevant thermodynamic quantities were calculated as functions of the nucleons density, temperature and  fractional exclusion statistics parameter $\alpha$. It has been shown that  at high temperatures and densities the thermodynamics of the system has a strong dependence on the statistics of the particles. The scenario in which the nucleon-nucleon interaction strength is independent of the statistics of particles was also calculated, but it leads in general to unstable thermodynamics.
\end{abstract}



\title{Fractional exclusion statistics applied to relativistic nuclear matter}

\maketitle

\section{Introduction}

The fractional exclusion statistics (FES), introduced by Haldane in Ref.~\cite{PhysRevLett.67.937.1991.Haldane}, received very much attention since its discovery and has been applied to many models of interacting particle systems and ideal gases in different external conditions (see for example Refs.~\cite{PhysRevLett.81.489.1998.Carmelo,PhysRevLett.72.600.1994.Veigy,JPhysB33.3895.2000.Bhaduri,PhysRevB.60.6517.1999.Murthy,NuclPhysB470.291.1996.Hansson,IntJModPhysA12.1895.1997.Isakov,PhysRevLett.73.3331.1994.Murthy,PhysRevLett.74.3912.1995.Sen,PhysRevLett.86.2930.2001.Hansson,PhysRevE.76.61112.2007.Potter,PhysRevE.75.61120.2007.Potter,JPA40.11255.2007.Comtet,JPA35.7255.2002.Anghel,RomRepPhys59.235.2007.Anghel,PhysLettA.372.5745.2008.Anghel,PhysRevLett.104.198901.2010.Anghel,EPL.94.60004.2011.Anghel,PhysRevE.76.061123.2007.Pellegrino}). Nevertheless, some of the basic properties of the model eventually have been deduced only recently \cite{EPL.87.60009.2009.Anghel,PhysRevLett.104.198901.2010.Anghel} and a general ansatz regarding the parameters of the FES have been introduced and applied in Refs. \cite{PhysLettA.372.5745.2008.Anghel,JPhysA.40.F1013.2007.Anghel,EPL.90.10006.2010.Anghel,RJP.54.281.2009.Anghel,JStatMech.P09011.2010.Anghel,EPL.94.60004.2011.Anghel}.

In Refs.~\cite{PhysLettA.372.5745.2008.Anghel,RJP.54.281.2009.Anghel} it was also shown that the FES is not an exceptional statistics, manifesting only in certain exotic systems, but it is generally present in interacting particle systems. Concretely, a system of interacting particles may be described as an ``ideal'' FES system, in which the exclusion statistics parameters are determined by the interaction Hamiltonian~\cite{PhysLettA.372.5745.2008.Anghel,RJP.54.281.2009.Anghel,JStatMech.P09011.2010.Anghel}.

In this paper we apply the formalism of FES to the relativistic nuclear matter described in the relativistic mean-field (RMF) model. The RMF is a general framework for the relativistic nuclear many-body systems based on hadronic degrees of freedom~\cite{AnnPhys.83.491.1974.Walecka,PhysLettB.79.10.1978.Serr,AdvNuclPhys.16.1.1986.Serot,IntJModPhys.6.515.1997.Serot} which has been successfully applied to describe many nuclear phenomena~\cite{PhysRep.409.101.2005.Vretenar}. The FES is applied to the description of the nucleons in RMF, to provide a general framework to study possible effects of the remnant interaction present in the model. This can yield useful predictions about the statistics of the particles and therefore of the effects of remnant interaction from the macroscopic properties of the system.

The structure of the article is the following. In the next section, we briefly describe basic ingredients of the relativistic mean-field model in fractional exclusion statistics approach and the methodology to evaluate the model parameters. The thermodynamic results are discussed in the third section. The main conclusions are summarized in the final section.

Throughout the paper we use the natural units system, $\hbar=c=k_B=1$.

\section{Implementation of FES into the relativistic mean-field model}\label{FES_QHD}
\subsection{The relativistic mean-field model}

Let us first introduce the notations and the basic concepts by following the treatment of Serot and Walecka from Ref.~\cite{AdvNuclPhys.16.1.1986.Serot}.

As already mentioned, RMF is a relativistic quantum theory model for the nuclear many-body system. The motivation to introduce the RMF was the observation of large Lorentz scalar and four-vector components in the nucleon-nucleon interaction.

In the model we have the baryonic field, $\psi$, that describes the nucleons, $p$ (protons) and $n$ (neutrons), the neutral scalar meson field, $\phi$ (for the meson $\sigma$) and the neutral vector meson field, $V_\mu$. It is assumed that the neutral scalar meson couples to the scalar density of baryons through $g_s\bar\psi\psi\phi$ and that the neutral vector meson couples to the conserved baryon current through $g_v\bar\psi\gamma_\mu\psi V^\mu$. The key approximation is to assume that the baryon density is high, so that we can use mean field theory and replace the meson field operators by their expectation values, which are the classical fields $\langle\phi\rangle\equiv\phi_0$ and $\langle V_\mu\rangle\equiv\delta_{\mu 0}V_0$. Then, for a static, uniform system, the quantities $\phi_0$ and $V_0$ are constants, independent of $x_\mu$. With these simplifications (see  Refs.~\cite{AdvNuclPhys.16.1.1986.Serot,IntJModPhys.6.515.1997.Serot} for details), we have $\phi_0 = \langle\bar\psi\psi\rangle g_s/m_s^2\equiv\rho_sg_s/m_s^2$ and $V_0=\langle\psi^\dagger\psi\rangle g_v/m_v^2\equiv\rho_B g_v/m_v^2$, with $m_s$ and $m_v$ the masses of the scalar and vector mesons, respectively, and the mean field Lagrangian is written as
\begin{eqnarray}
\cL &=& \bar\psi\left[i\gamma_\mu\partial^\mu - g_v\gamma^0V_0 -
(M-g_s\phi_0)\right]\psi - \frac{1}{2}m_s^2\phi_0^2 +
\frac{1}{2}m_v^2V_0^2 . \label{cLMFT}
\end{eqnarray}
The initial Lagrangian of the RMF model and more detailed description of the model can be found in Refs.~\cite{AdvNuclPhys.16.1.1986.Serot,IntJModPhys.6.515.1997.Serot}. Applying the Lagrange equations to $\cL$ and solving for plane-waves, we obtain the quasiparticle energies, $\epsilon(k)=E^*(k)+g_vV_0$ and $\bar\epsilon(k)=E^*(k)-g_v V_0$, with $E^*(k)=\sqrt{\bk^2+M^{*2}}$ and $M^*= M-g_s\phi_0$ the effective mass. The baryonic number, $\hat{B}\equiv\int_Vd^3\bx \hat{B}^0 =\int_Vd^3\bx\psi^\dagger\psi$, is a constant of motion, so it is conserved. (Throughout the paper, we shall denote the volume of the hadronic system by $V$--without any index.)

From the energy-momentum tensor calculated with (\ref{cLMFT}) we derive the total Hamiltonian of the system and the expressions for the pressure and energy density. In the second quantization, the baryon number and the effective Hamiltonian operators are
\begin{eqnarray}
\hat B &=& \sum_{\bk,\lambda}[A^\dagger_{\bk\lambda}A_{\bk\lambda}-
B^\dagger_{\bk\lambda}B_{\bk\lambda}],
\label{BN} \\
\hat H &=& \sum_{\bk,\lambda} E^*(k)
[A^\dagger_{\bk\lambda}A_{\bk\lambda} +
B^\dagger_{\bk\lambda}B_{\bk\lambda}]
 +g_v V_0\hat B \nonumber  \\ &&
-V\left(\frac{1}{2}m_v^2V_0^2-\frac{1}{2}m_s^2\phi_0^2\right), \label{Heff2q}
\end{eqnarray}
where $A^\dagger_{\bk\lambda}$ and $A_{\bk\lambda}$ are the creation and annihilation operators for the baryon state of momentum $\bk$ and species $\lambda$ ($\lambda=n$ or $p$, of spin up or spin down), whereas $B^\dagger_{\bk\lambda}$ and $B_{\bk\lambda}$ are the corresponding antiparticle operators.

We assume that all the protons and neutrons have the same mass, and therefore their spectra are identical.

\subsection{FES implementation}

Into this formalism we introduce the FES. We shall consider FES in its simplest form, in which we have a single, direct exclusion  statistics parameter $\alpha$. We introduce the spin-isospin degeneracy factor, $\gamma$, which takes the value 2 for neutron matter (two spin projections and one isospin projection) and 4 for nuclear matter (two spin and two isospin projections). The net baryon number $B$ is a conserved quantity, so the chemical potential, $\mu$, is associated to it.

In the grand canonical ensemble, at temperature $T$ and chemical potential $\mu$, the thermodynamic potential and the partition function are
\begin{equation}\label{cZ}
\Omega=-T\ln \cZ \qquad {\rm and} \qquad \cZ = Tr\left(e^{-\frac{\hat H -\mu \hat B}{T}}\right),
\end{equation}
respectively. Plugging Eqs. (\ref{BN}) and (\ref{Heff2q}) into (\ref{cZ}) we obtain in the standard way~\cite{AdvNuclPhys.16.1.1986.Serot,IntJModPhys.6.515.1997.Serot} the baryonic density, $\rho_B$, the energy density, $\cE$, and the pressure, $p$,
\begin{subequations}\label{Eq_state_T}
\begin{equation}
\rho_B\equiv \frac{\langle \hat{B}\rangle}{V}= -\frac{1}{V}\left(\frac{\partial\Omega}{\partial\mu}\right)_{TV} =
\frac{\gamma}{(2\pi)^3}\int_{R^3} d^3\bk [n_{k}(T)-\bar n_{k}(T)], \label{Eq_state_T_rho}
\end{equation}
\begin{eqnarray}
\cE &\equiv& \frac{\langle \hat{H}\rangle}{V} = -\frac{T^{2}}{V}\frac{\partial}{\partial T}\left(\frac{\Omega}{T}\right)_{V\mu}+\mu\rho_{B}  = \frac{g_v^2}{2m_v^2}\rho_B^2 +\frac{m_s^2}{2g_s^2}(M-M^*)^2 \nonumber \\ && +
\frac{\gamma}{(2\pi)^3}\int_{R^3}d^3\bk E^*(k) [n_{k}(T)+\bar n_{k}(T)], \label{Eq_state_T_E} \\
p &=&-\left(\frac{\partial\Omega}{\partial V}\right)_{T\mu} = \frac{g_v^2}{2m_v^2}\rho_B^2 - \frac{m_s^2}{2g_s^2}(M-M^*)^2 \nonumber \\ &&
+ \frac{\gamma}{3(2\pi)^3}\int_{R^3}d^3\bk\frac{\bk^2}{E^*(k)}
[n_{k}(T)+\bar n_{k}(T)], \label{Eq_state_T_p}
\end{eqnarray}
where $n_{k}(T)$ and $\bar n_{k}(T)$ are the nucleon and antinucleon fermionic mean occupation numbers, 
\begin{equation}
n_k(T)=\frac{1}{e^{(E^*(k)-\nu)/T}+1}, \qquad
\bar n_k(T)=\frac{1}{e^{(E^*(k)+\nu)/T}+1}, \label{nnbarF}
\end{equation}
with $\nu\equiv\mu-g_vV_0=\mu-g_v^2\rho_B/m_v^2$. 

The equilibrium values of the fields $\phi_0$ and $V_0$, and implicitly that of $M^*$, are determined by the minimization of the thermodynamical potential with respect to $\phi_{0}$ and $V_{0}$, $\partial\Omega/\partial \phi_{0}=0$ and $\partial\Omega/\partial V_{0}=0$, respectively, which give the equations \cite{AdvNuclPhys.16.1.1986.Serot,IntJModPhys.6.515.1997.Serot},
\begin{eqnarray}
M-M^*-\frac{g_s^2}{m_s^2}\frac{\gamma}{(2\pi)^3}\int_{R^3}d^3\bk
\frac{M^*[n_{k}(T)+\bar n_{k}(T)]}{\sqrt{\bk^2+M^{*2}}} &=& 0,
\label{T_mass} \\
V_{0}-\frac{g_{v}}{m_{v}^{2}} \rho_{B} &=& 0; \label{V_mass}
\end{eqnarray}
Eq.~(\ref{T_mass}) has to be solved self-consistently for $M^*$.

The minimization of the thermodynamic potential with respect to the mean-fields provides self-consistent thermodynamic relations in the variables of state $(T,V,\mu)$: $(\partial\Omega/\partial T)_{V\mu}\equiv(\partial\Omega/\partial T)_{V\mu \phi_{0}V_{0}}$, $(\partial\Omega/\partial V)_{T\mu}\equiv(\partial\Omega/\partial V)_{T\mu \phi_{0}V_{0}}$, and $(\partial\Omega/\partial \mu)_{TV}\equiv(\partial\Omega/\partial \mu)_{TV \phi_{0}V_{0}}$.
\end{subequations}

The effect of introducing FES in the formalism, through Eq. (\ref{cZ}),  is that the mean occupation numbers are not anymore given by Eq.~(\ref{nnbarF}). The baryon and antibaryon spectra are coarse-grained and each grain represents a FES species. Each species contains $N_{E^*,i}$ ($\bar N_{E^*,i}$) particles (anti-particles) and has the dimension $G_{E^*,i}$, which is defined as the number of single-particle states in the grain before putting in the particles (anti-particles); by $E^*$ and $i$ we identify uniquely the grain. The $N_{E^*,i}$ particles change the number of available states in the species by $-\alpha N_{E^*,i}$ and this leads to a number of $W_{E^*,i}=(G_{E^*,i}+(1-\alpha)N_{E^*,i}-1+\alpha)! /[(G_{E^*,i}-\alpha N_{E^*,i}-1+\alpha)! N_{E^*,i}!]$ microconfigurations in which these particles may be arranged. (Similar expressions may be written for anti-particles.) Writing the number of microconfigurations for each grain, plugging them into Eq. (\ref{cZ}) and maximizing (minimizing) $\cZ$ ($\Omega$) with respect to $n_{\bk,\alpha}\equiv N_{E^*(\bk),i}/G_{E^*(\bk),i}$ and $\bar n_{\bk,\alpha}\equiv\bar N_{E^*(\bk),i}/G_{E^*(\bk),i}$, we obtain the particle and anti-particle populations ~\cite{JPhysA.40.F1013.2007.Anghel,PhysRevLett.73.2150.1994.Isakov,PhysRevLett.73.922.1994.Wu}
\begin{equation}
n_{k,\alpha}(T) = [w+\alpha]^{-1}, \qquad \bar n_{k,\alpha}(T)= [\bar w+\alpha]^{-1},  \label{Eqnw}
\end{equation}
where $w$ and $\bar w$ are the solutions of the system
\begin{equation}
w^\alpha[1+w]^{1-\alpha}=\xi, \qquad \bar w^\alpha[1+\bar w]^{1-\alpha}=\bar \xi, \label{Eqw}
\end{equation}
with $\xi= e^{(E^*(k)-\nu)/T}$ and $\bar \xi= e^{(E^*(k)+\nu)/T}$ (we added to $n$ the subscript $\alpha$ to specify the statistics). One can notice by inspection that $\alpha=0$ and $\alpha=1$ reproduce the Bose and Fermi statistics, respectively.

Therefore the formalism remains the same, except that we work with the more general level populations (\ref{Eqnw}), instead of (\ref{nnbarF}).

This generalized model contains three phenomenological parameters, $C_s^2\equiv g_s^2M^2/m_s^2$, $C_v^2\equiv g_v^2M^2/m_v^2$ and $\alpha$. We determine $C_s$ and $C_v$ from the physical conditions for the nuclear matter at the saturation point~\cite{AdvNuclPhys.16.1.1986.Serot}--at $T=0$ and $\rho_B\equiv\rho_{B0}$, the binding energy, $\cE_b\equiv[\cE(\rho_B)-\rho_BM]/\rho_B$, attains its minimum (ground state) value, $\cE_{b0}$--whereas $\alpha$ is varied freely. For the nuclear matter we take~\cite{IntJModPhys.6.515.1997.Serot}
\begin{equation}
\cE_{b0}\approx -16\ {\rm MeV}, \qquad  \rho_{B0} = 0.16 \ {\rm fm}^{-3}. \label{Eb0rhoB0}
\end{equation}
Therefore we rewrite Eqs.~(\ref{Eq_state_T}) at $T=0$:
%
\begin{equation}
\rho_B = \frac{\gamma}{(2\pi)^3\alpha}\int_{k<k_F}d^3\bk
=\frac{\gamma}{6\pi^2\alpha}k_F^3 , \label{Eq_state_T0_rho}
\end{equation}
and
\begin{subequations}\label{Eq_state_T0}
\begin{eqnarray}
\cE &=& \frac{g_v^2}{2m_v^2}\rho_B^2 +
\frac{m_s^2}{2g_s^2}(M-M^*)^2 +
\frac{\gamma}{(2\pi)^3\alpha}\int_{k<k_F}d^3\bk E^*(k) ,
\label{Eq_state_T0_E} \\
p &=& \frac{g_v^2}{2m_v^2}\rho_B^2 - \frac{m_s^2}{2g_s^2}(M-M^*)^2
+\frac{\gamma}{3(2\pi)^3\alpha}\int_{k<k_F}d^3\bk\frac{\bk^2}{E^*(k)},
\label{Eq_state_T0_p} \\
0 &=& M-M^*-\frac{g_s^2}{m_s^2}\frac{\gamma}{(2\pi)^3\alpha}\int_{k<k_F}d^3\bk
\frac{M^*}{E^*(k)} , \label{T0_mass}
\end{eqnarray}
\end{subequations}
where $k_F$ is the Fermi wave-vector. If we minimize $\cE_b$ for the usual Fermi statistics, $\alpha=1$, and with the expression for $\cE$ given by (\ref{Eq_state_T0_E}), we obtain $C_s^2\approx330$ and $C_v^2\approx249$.

\section{Thermodynamics}

\begin{figure}[t]
\begin{center}
\includegraphics[width=14cm]{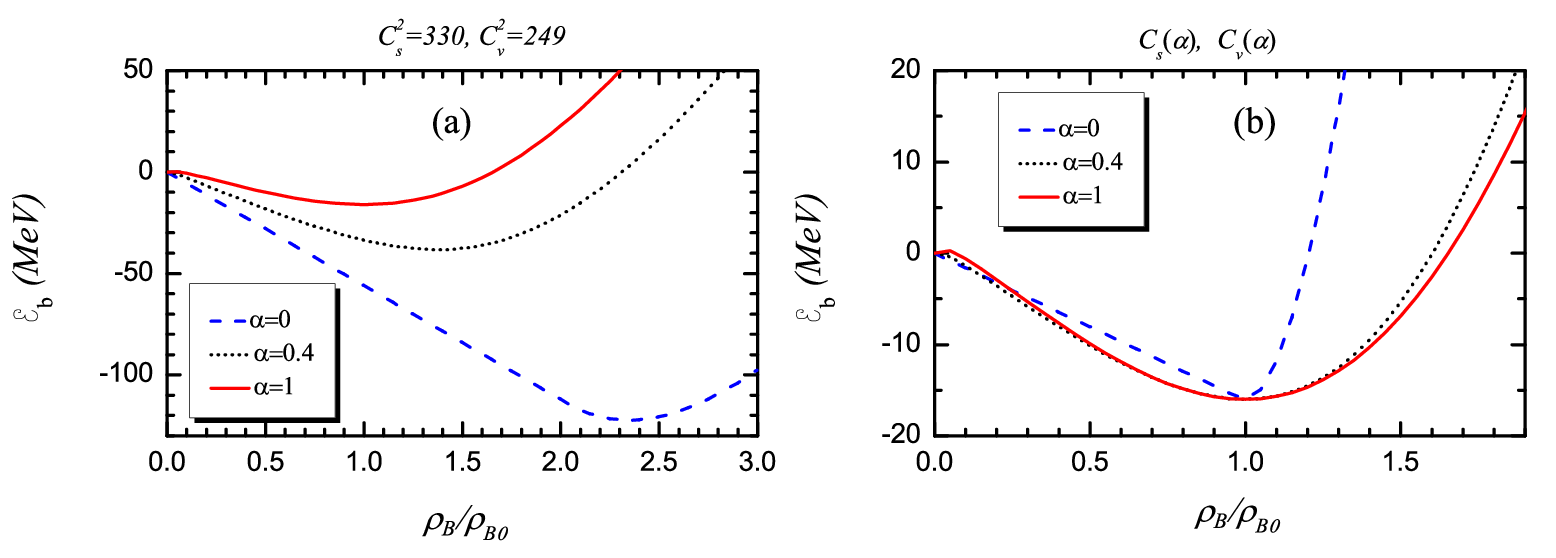} 
\caption{(Color online) The binding energy per nucleon, $\cE_{b}$, as a function of the nucleons density $\rho_{B}$ for the isospin-symmetric nuclear matter at zero temperature, for various values of the parameter $\alpha$. (a) The curves were calculated with fixed coupling constants, determined from the saturation point in Fermi statistics ($\alpha=1$), $C_s^2=330$ and $C_v^2=249$, or (b) with running coupling constants, $C_s$ and $C_v$ as functions of $\alpha$, determined from the condition that the saturation point is the same for any $\alpha$: $\cE_{b}=-16$ MeV and $\rho_{B}=0.16$ fm$^{-3}$.} \label{E_B_T0_2in1}
\end{center}
\end{figure}

Now we can calculate the properties of the system at finite temperatures. We can imagine two scenarios. In the first scenario the parameters $C_s$ and $C_v$ are fixed at $C_s^2=330$ and $C_v^2=249$ for any $\alpha$ and therefore the saturation point varies with $\alpha$ (Fig. \ref{E_B_T0_2in1} a), whereas in the second scenario the saturation point is always located at $\cE_{b0}$ and $\rho_{B0}$ and therefore $C_s$ and $C_v$ depend on $\alpha$ (Fig. \ref{E_B_T0_2in1} b).

\subsection{Scenario 1: $C_s$ and $C_v$ are fixed}\label{scenario1}

Let us analyze scenario 1, with $C_s$ and $C_v$ fixed by the condition at the saturation point for $\alpha=1$--$C_s^2=330$ and $C_v^2=249$. For the clarity of the calculations, we introduce the scaled variables $r\equiv M^*/M$ and $x\equiv M/k_F$. In these notations, the Eqs. (\ref{Eq_state_T0}) become
\begin{subequations}\label{Eq_T0_anrx}
\begin{eqnarray}
\cE &=& \frac{C_v\rho_B^2 }{2M^2} +\frac{(2-r)(1-r)M^4}{4C_s} +\frac{3}{4} \rho_B\sqrt{k_F^2+r^2M^2}
\label{T0_E_anrx} \\
p &=& \frac{C_v\rho_B^2 }{2M^2} -\frac{(2-r)(1-r)M^4}{4C_s} +\frac{1}{4} \rho_B\sqrt{k_F^2+r^2M^2}
\label{T0_p_anrx} \\
0 &=& 1-r-\frac{3}{2}\frac{C_s\rho_B}{M^3}
rx\left[\sqrt{1+r^2x^2}-x^2r^2\ln\frac{1+\sqrt{1+r^2x^2}}{rx}\right].
\label{Eqrx2}
\end{eqnarray}
\end{subequations}
We observe that in Eqs. (\ref{Eq_T0_anrx}) $\alpha$ enters only indirectly, through $k_F$ and $x$. When $\alpha$ decreases to zero, so does $k_F$ and, as a consequence, $x$ diverges to infinity.

\begin{figure}[t]
  \begin{center}
  \includegraphics[width=7cm]{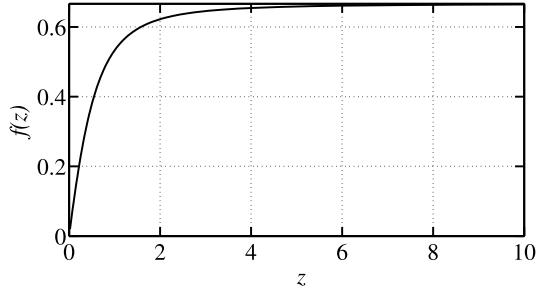}
\end{center}
\caption{The function $f$, of Eq. (\ref{fdez}).} \label{fdez_plot}
\end{figure}

Let us analyze the solutions of the mass equation (\ref{Eqrx2}). For this, we define the function
\begin{equation}
  f(z) \equiv z\left[\sqrt{1+z^2}-z^2\ln\frac{1+\sqrt{1+z^2}}{z}\right]
  \label{fdez}
\end{equation}
and rewrite Eq. (\ref{Eqrx2}) as
\begin{equation}
  1-r = \frac{3}{2}\frac{C_s\rho_B}{M^3} f(rx). \label{Eqrx2f}
\end{equation}
Without going deep into the analysis of Eq. (\ref{Eqrx2f}), we observe that $f(z)$ is a monotonically increasing function, with $f(0)=0$ and $f(\infty)=2/3$ (see Fig. \ref{fdez_plot}). Since, on the other hand, $1-r$ decreases monotonically from 1 to 0, as $r$ increases from 0 to 1, and since $f(x)>0$, then Eq. (\ref{Eqrx2f}) always has a unique solution, $r\in(0,1)$, for any finite $x$.

Now let us see what happens when $\alpha\to0$. In this case $x\to\infty$ and we have two possibilities: (1) if $r$ remains finite, then $rx\to\infty$, and (2) if $r\searrow0$, then $rx$ may converge to a finite value that we shall call $\overline{rx_0}$.

In the \textit{case (1)}, since $\lim_{z\to\infty}f(z)=2/3$, Eq. (\ref{Eqrx2f}) gives the simple solution,
\begin{equation}
  r_{\alpha=0} = 1-\frac{M^3}{C_s\rho_B}, \label{r_case1}
\end{equation}
which may be true only if $r>0$, and therefore as long as $M^3/(C_s\rho_B)<1$.

If $M^3/(C_s\rho_B)<1$ (case 1) is not satisfied, we are in \textit{case (2)}, in which $r=0$ and $\overline{rx_0}$ is determined by the equation
\begin{equation}
  \frac{3}{2}\frac{C_s\rho_B}{M^3} f(\overline{rx_0})=1. \label{r_case2}
\end{equation}
Equation (\ref{r_case2}) has a solution $\overline{rx_0}>0$ for any $M^3/(C_s\rho_B)\ge1$.

We observe here an interesting case of symmetry breaking when $\rho_B=\rho_{B0}>0$ while $M^*=0$. Due to the particle-antiparticle symmetry, in a system of bosons of zero rest mass the chemical potential, $\nu$, should be zero at any temperature. This implies that the particle and antiparticle excited energy levels are always equally populated and therefore a nucleons density different from zero, like in case (2), can be realized only by an asymmetric and macroscopic population of the particle and antiparticle ground-states.

\begin{figure}[t]
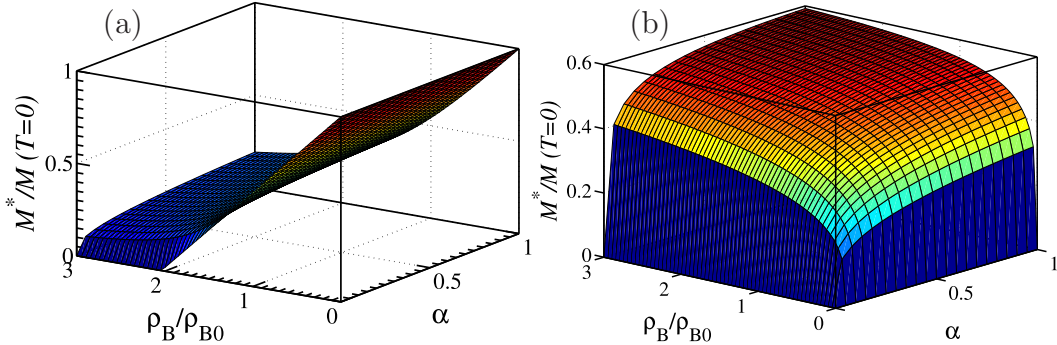

  \begin{center}
    \unitlength1mm\begin{picture}(140,45)
    \put(0,0){\includegraphics[width=7cm]{m0_CsCv_fix_surface.epsi}}
    \put(70,0){\includegraphics[width=7cm]{m0_CsCv_alpha_surface.epsi}}
    \put(13,41){(a)}
    \put(83,41){(b)}
    \end{picture}
  \end{center}
  \caption{(Color online) The ratio $r=M^*/M$ vs. $\alpha$ and $\rho_B$ for scenario (1), with $C_s^2=330$ and $C_v^2=249$, fixed (a), and for scenario (2), with the saturation point fixed and $C_s$ and $C_v$ dependent on $\alpha$ (b). We observe that in (a) $r(\alpha=0)>0$ for $\rho_B<M^3/C_s\approx2.04$.}
  \label{m0_CsCv}
\end{figure}

The two cases, $M^3/(C_s\rho_B)\le1$ and $M^3/(C_s\rho_B)>1$, appear clearly in Fig. \ref{m0_CsCv} (a), where $r_{\alpha=0}>0$ for small values of $\rho_B$ and $r_{\alpha=0}=0$ for larger $\rho_B$. The limit between the two cases is $M^3/C_s$, which, for the parameters we use in this paper is approximately $2.04$.

\begin{figure}
  \begin{center}
\includegraphics[width=14cm]{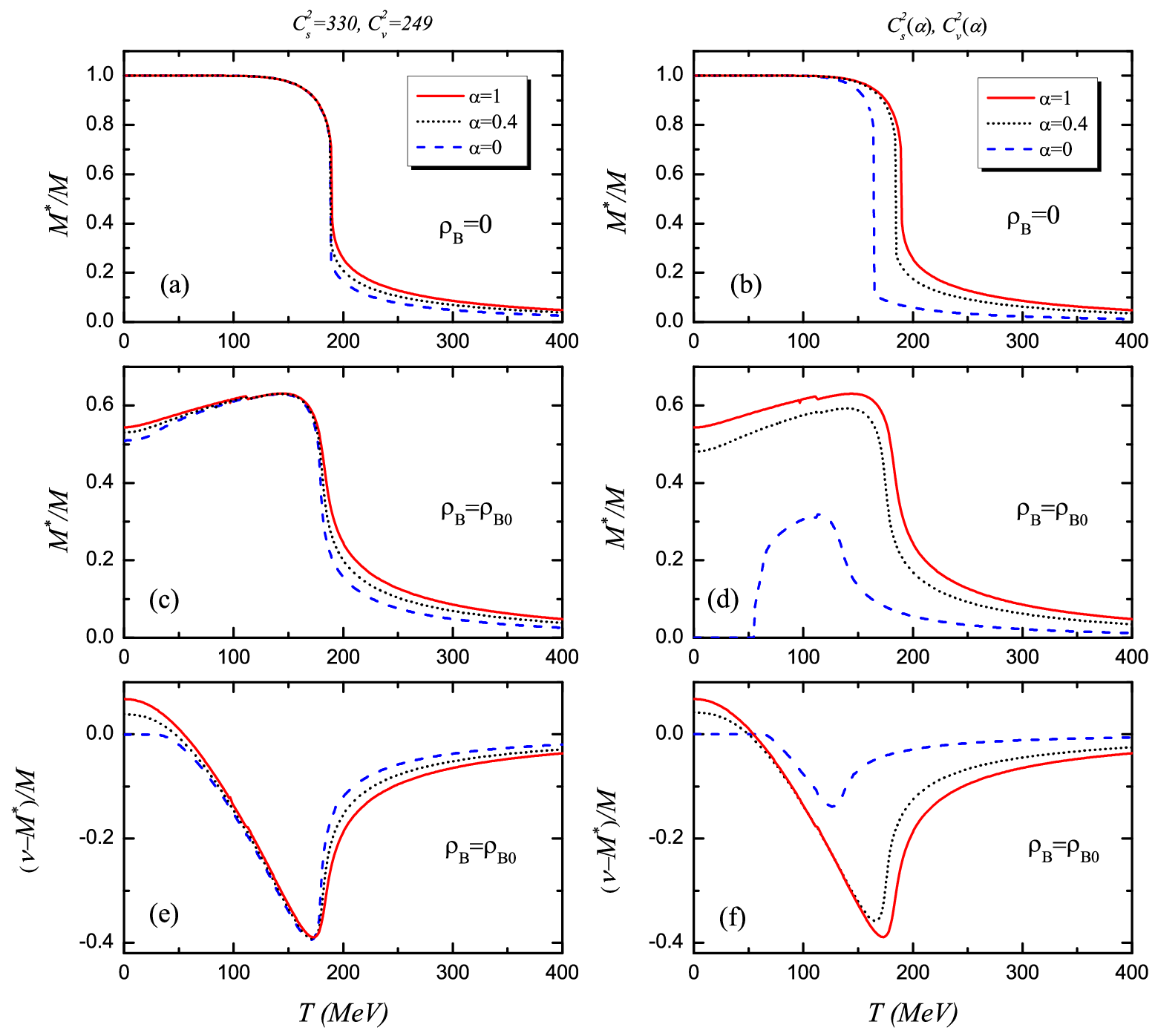} 
  \end{center}
  \caption{(Color online) The effective mass, $M^*$ and the relative chemical potential, $(\nu-M^*)/M$, both, normalized to the free nucleon mass, $M$, vs. $T$. All plots correspond to nuclear matter, $\gamma=4$. In the left column we plot the results obtained in scenario (1), with $C_s\approx330$ and $C_v\approx249$ being fixed, while in the right column we have the results of scenario 2, when the saturation point is the same for all values of $\alpha$, which renders coupling constants, $C_s$ and $C_v$, dependent on $\alpha$.}
  \label{M_nu_g4}
\end{figure}

The plots on the left column of Fig.~\ref{M_nu_g4} correspond to $C_s^2=330$ and $C_v^2=249$, fixed. In the plots (a) and (c) we show the effective mass of the nucleons for $\rho_B=0$ and $\rho_B=\rho_{B0}$, respectively. We observe that if $\rho_B=0$, $M^*$ converges to $M$ at $T\to0$, for any $\alpha$. This result is immediately obtained if we set $\rho_B=0$ in Eq.~(\ref{Eqrx2}) and can also be seen in Fig.~\ref{m0_CsCv} (a). For $\rho_B=\rho_{B0}$, the effective mass converges to a finite value smaller than $M$, when $T\to0$ as we discussed above. This finite value depends on the statistics.

In Fig. \ref{M_nu_g4} (e) we plot the normalized relative chemical potential, $(\nu-M^*)/M$. Here we observe a phenomenon which is specific to this system. In general, if the single-particle density of states of a system increases with the particle energy, the chemical potential is expected to decrease monotonically with temperature. In our system, although $\nu-M^*$ decreases with $T$ at low temperatures and becomes negative, if we increase the temperature further, at a temperature between 150 and 200 MeV it has an upwards turn and then increases monotonically to zero. This phenomenon is associated to the decrease of $M^*$ with $T$, in the high temperature range. As the temperature increases, the mass of the particle decreases to zero and the (relative) chemical potential also increase from negative values, to zero, since a gas of massless particles can have only zero chemical potential.

Also in Fig.~\ref{M_nu_g4} (e) we observe that for $\alpha=0$, at a temperature slightly below 30~MeV, the relative chemical potential, $\nu-M^*$, becomes zero, signaling a Bose-Einstein condensation (BEC).

\begin{figure}
\begin{center}
\includegraphics[width=13cm]{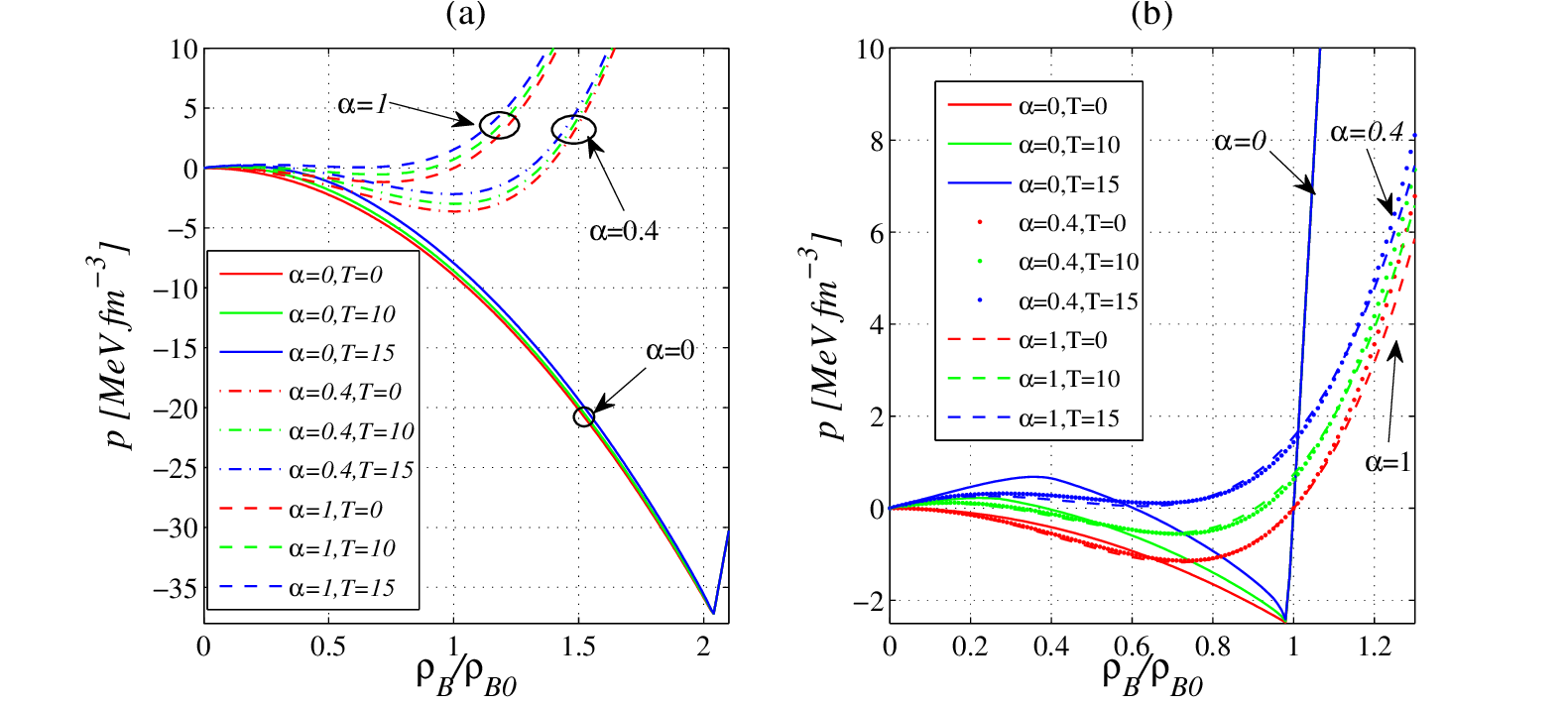}
\caption{(Color online) Isotherms of pressure versus nucleons density in the generalized RMF approximation for the nuclear matter at finite temperature $T$ for various values of the parameter $\alpha$. In (a) we plot the results obtained in scenario (1), with $C_s\approx330$ and $C_v\approx249$ being fixed, while in (a) we plot the results of scenario 2, when the saturation point is the same for all values of $\alpha$, which renders coupling constants, $C_s$ and $C_v$, dependent on $\alpha$.} \label{Isotherms}
\end{center}
\end{figure}

In Fig.~\ref{Isotherms} we plot the isotherms of the nuclear matter at three different temperatures, $T=0,10$, and 15 MeV, each temperature for three values of $\alpha$, namely $\alpha=0,0.4$, and 1. The results for the scenario (1) are plotted in Fig.~\ref{Isotherms} (a). For each value of $\alpha$, the isotherms of higher temperatures lie above those of lower temperature--the system expands when heated. We also observe that the isotherms for $\alpha=0$ have a very sharp turn upwards at a density $\rho_B$ slightly above $2\rho_{B0}$. Nevertheless, eventually one of the most important things to notice in Fig.~\ref{Isotherms} (a) is that for small $\alpha$'s the pressure of the gas is negative and therefore the system is unstable.

We turn now to the scenario (2), in which the system is stable for any $\alpha$.

\subsection{Scenario 2: $C_s$ and $C_v$ functions of $\alpha$}\label{scenario2}

In the scenario 2 we determine the parameters $C_s$ and $C_v$ as functions of $\alpha$ by fixing the minimum of the binding energy at the experimentally observable value, $\cE_b(\rho_{B0})=\cE_{b0}$. The condition of minimum for $\cE_b$ reads
\begin{eqnarray}
\frac{d}{d\rho_B}\left(\frac{\cE-\rho_BM}{\rho_B}\right)_{\rho_{B0}}
&=& \frac{1}{\rho_{B0}}\left.\frac{d\cE}{d\rho_B}\right|_{\rho_{B0}}
-\frac{\cE(\rho_{B0})}{\rho_{B0}^2} = 0 \label{eq_cond}
\end{eqnarray}
and calculating the total derivative of $\cE$ with respect to $\rho_B$ from Eq.~(\ref{Eq_state_T0_E}), we obtain an equation for $C_v$,
\begin{equation}
\frac{C_v^2}{M^2} = \frac{\cE_{b0}+M-E^*(k_{F0})}{\rho_{B0}} ,
\label{EqCv}
\end{equation}
where $k_{F0}\equiv [(2\pi)^3\alpha \rho_{B0}/\gamma]^{1/3}$. Using Eq. (\ref{EqCv}) we eliminate $C_v$ from the expression of $\cE$ (\ref{Eq_state_T0_E}),
\begin{eqnarray}
\cE &=& \rho_{B}\frac{\cE_{b0}+M-E^*(k_{F0})}{2}
+ \frac{M^2(M-M^*)^2}{2C_s} \nonumber \\
&& + \frac{\gamma}{16\pi^2\alpha}
\left[k_FE^*(k_{F0})(2k_F^2+M^{*2}) - M^{*4}
\ln\frac{k_{F0}+E^*(k_{F0})}{M^*}\right], \label{EqCs}
\end{eqnarray}
where $k_F$ is given by Eq. (\ref{Eq_state_T0_rho}) as a function of $\rho_B$ and we evaluate analytically the integral over $\bk$. From the effective mass equation (\ref{T0_mass}), we get $C_s$ as a function of $M^*$,
\begin{equation}
C_s = \frac{4\pi^2\alpha}{\gamma}\cdot\frac{M^2(M-M^*)}{M^*}\cdot
\left[k_F\sqrt{k_F^2+M^{*2}} - M^{*2}
\ln\frac{k_F+\sqrt{k_F^2+M^{*2}}}{M^*}\right]^{-1} \label{formCs}
\end{equation}
which we insert into (\ref{EqCs}) to obtain a self-consistent equation for $M^*$:
\begin{eqnarray}
\frac{\cE_0}{2} &=& -\frac{\rho_{B0}E^*(k_{F0})}{2} + \frac{\gamma
M^{*}(M-M^*)}{8\pi^2\alpha}\left\{
k_{F0}E^*(k_{F0})\right. \nonumber\\
&&\left.-M^{*2}\ln\frac{k_{F0}+E^*(k_{F0})}{M^*}\right\}
+ \frac{\gamma}{16\pi^2\alpha}
\left[k_{F0}E^*(k_{F0})(2k_{F0}^2+M^{*2}) \right. \nonumber \\
&& \left.- M^{*4}\ln\frac{k_{F0}+E^*(k_{F0})}{M^*}\right]. \label{EqCs2}
\end{eqnarray}
With $M^*$ calculated from the equation above, we go back and calculate $C_s$ and $C_v$ from Eqs.~(\ref{formCs}) and (\ref{EqCv}).

To study the solutions of Eq. (\ref{EqCs2}) we write it in terms of the dimensionless parameters and function $f$ (\ref{fdez}),
\begin{equation}
\frac{4\cE}{\rho_B}\equiv 4M\left(1+\frac{\cE_{b0}}{M}\right) = 2\sqrt{y_F^2+r^2M^2}+3M(2-r)f(rx).
\label{EqCs3}
\end{equation}
For the limit $\alpha\to0$ we have again the two situations from section \ref{scenario1}: (1) $\lim_{\alpha\to0}r>0$, so $\lim_{\alpha\to0}rx=\infty$, and (2) $\lim_{\alpha\to0}r=0$, so that $\lim_{\alpha\to0}rx$ is finite. But in case (1), Eq. (\ref{EqCs3}) becomes
\begin{equation}
4M\left(1+\frac{\cE_{b0}}{M}\right) = 4M, \label{EqMalpha0}
\end{equation}
which is a contradiction, since $\cE_{b0}<0$. Therefore the only possibility is that $\lim_{\alpha\to0}r=0$ and $\lim_{\alpha\to0}rx=\overline{rx_0}<\infty$. In Fig.~\ref{m0_CsCv} (b) we plot the relative mass, $r=M^*/M$, as a function of $\alpha$ and the relative density, $\rho_B/\rho_{B0}$ corresponding to this scenario and we observe that indeed, $r=0$ at $\alpha=0$ for any $\rho_B$.

The value of the product $rx=\overline{rx_0}$, at $\alpha=0$ may be obtained from the equation
\begin{eqnarray}
4M\left(1+\frac{\cE_{b0}}{M}\right) &=& 6M\overline{rx_0}\left[\sqrt{1+\overline{rx_0}^2} - \overline{rx_0}^2\ln\frac{1+\sqrt{1+\overline{rx_0}^2}}{\overline{rx_0}}\right] \nonumber\\
&\equiv& 6Mf(\overline{rx_0}),\label{EqCs3a0}
\end{eqnarray}
obtained from  Eq. (\ref{EqCs3}).

Plugging the solution of Eq.~(\ref{EqCs3}) into Eqs.~(\ref{EqCv}) and (\ref{formCs}), we obtain the limits of $C_v$ and $C_s$ at $\alpha=0$, namely
\begin{equation}
C_v^2(\alpha=0) = \frac{M^2\cE_0}{\rho_B^2} \quad {\rm and} \quad C^2_s(\alpha=0) = \frac{M^4}{\cE_{0}}.
\label{limCsa0}
\end{equation}
At $\alpha=1$ and $\rho_B=\rho_{B0}$ we obtain the typical results, $C_s^2(1)\approx330$ and $C_v^2(1)\approx249$.

\begin{figure}
\begin{center}
\includegraphics[width=14cm]{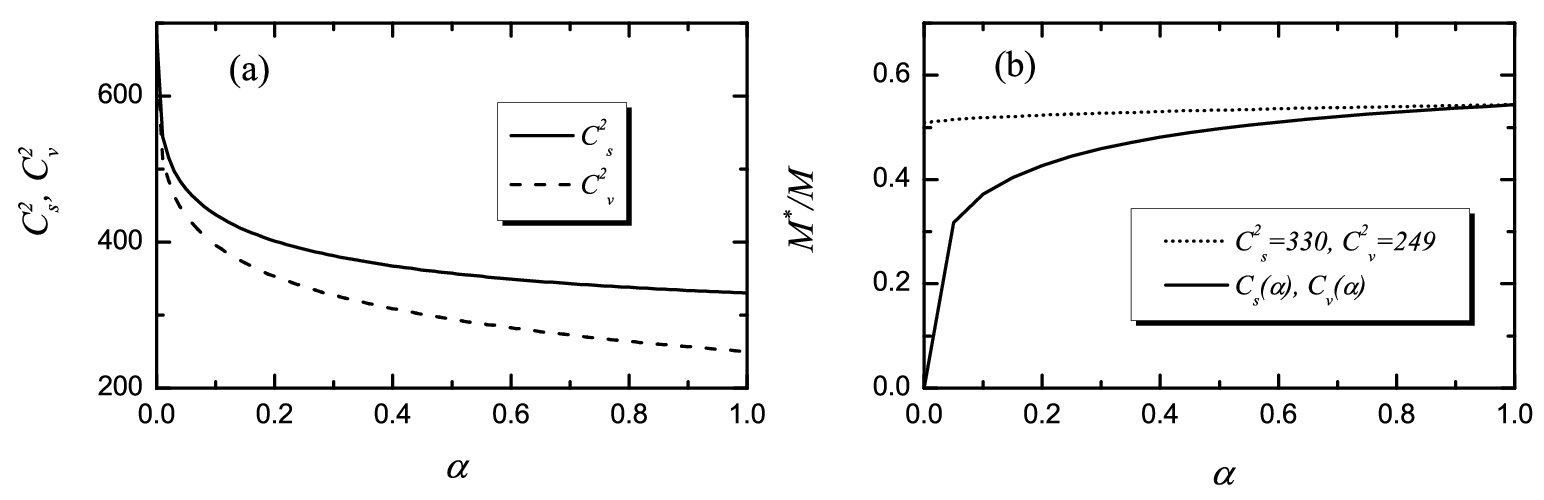}\vspace{-0.4cm}
\caption{(a) The dependence of the parameters $C_s$ and $C_v$ on the exclusion statistics parameter $\alpha$ for the generalized RMF model in the scenario (2). (b) The effective nucleon mass vs. $\alpha$ in both, the scenario (1) (dotted line) and scenario (2) (solid line). } \label{Cs_Cv_Meff_alpha}
\end{center}
\end{figure}

The parameters $C_s^2$ and $C_v^2$ are plotted in Fig.~\ref{Cs_Cv_Meff_alpha} (a) as functions of $\alpha$, for $\rho_B=\rho_{B0}$. In Fig.~\ref{Cs_Cv_Meff_alpha} (b) we plot the relative mass, $r$, as a function of $\alpha$, also for $\rho_B=\rho_{B0}$, corresponding to the scenario 1 (dot line) and to scenario 2 (solid line). The two lines are cross-sections through the three-dimensional plots of Fig.~\ref{m0_CsCv}.

Equation (\ref{EqCs3}) may be put into the dimensionless form
\begin{equation}
  4x\left(1+\frac{\cE_{b0}}{M}\right) = 2\sqrt{1+r^2x^2}-3rxf(rx)+6xf(rx). \label{EqCs3age0}
\end{equation}
Calculating $r$ from Eq.~(\ref{EqCs3age0}) and then $C_s$ and $C_v$ for each $\alpha$, we are able to solve the finite temperature mass equation (\ref{T_mass}). The results are plotted in Figs.~\ref{M_nu_g4} (b), for $\rho_B=0$, and (d), for $\rho_B=\rho_{B0}$. We observe that the dependence of $M^*$ on $T$ for $\rho_B=0$ is quite similar in the two scenarios (Figs.~\ref{M_nu_g4} a and b).

The qualitative difference between the results of the two scenarios appear at $\rho_B>0$. We observe in Fig.~\ref{M_nu_g4} (d) that for $\alpha=0$, $M^*$ becomes zero at finite temperatures, while in the first scenario $M^*$ remains positive at any temperature for $\rho_B=\rho_{B0}$.

To elucidate the reason for the mass sudden disappearance in the scenario (2) at finite temperatures, we plotted in Fig.~\ref{M_nu_g4} (f) the relative chemical potential, $(\nu-M^*)/M$. In this way we observe that the gas of $\alpha=0$ undergoes a BEC--like in the scenario (1)--but at a temperature which is slightly higher than the temperature at which $M^*$ becomes zero. Therefore the sudden mass disappearance and the BEC are not directly related, although they may influence one-another. Moreover, the BEC appears in the scenario (2) at higher temperatures than in the scenario (1).

In Fig.~\ref{Isotherms} (b) we plot the isotherms and we observe that at $\rho_B=\rho_{B0}$ the system is the stable ($p>0$) at any $\alpha$, unlike in the scenario (1) (Fig.~\ref{Isotherms} a), when the system becomes  unstable for small $\alpha$'s. Moreover, at low temperatures and nucleons densities, the isotherms depend very little on the exclusion statistics parameter, whereas at high temperatures the isotherms become very sensitive to $\alpha$. This is due to the fact that the phenomenological constants $C_s$ and $C_v$ are recalculated for each $\alpha$, so that the saturation point is independent of the exclusion statistics.

On the contrary, if $C_s$ and $C_v$ are the same for all $\alpha$'s, then the Hamiltonian of the system is independent of $\alpha$ and the high temperature results coincide (statistics has smaller and smaller influence on the thermodynamic results as the temperature increases), whereas at low temperatures the isotherms strongly depend on $\alpha$~\cite{ISHEPP.1.167.2008.Anghel}, as expected from the thermodynamics of the ideal quantum gas.

\section{Conclusions}\label{concl}

In the present paper we analyzed the effect of the statistics of particles on the thermodynamic properties of the nuclear matter. For this, we generalized the relativistic mean-field model to include fractional exclusion statistics. The generalized RMF model is thermodynamically self-consistent. The parameters of the model are $\alpha$ (the exclusion statistics parameter), $C_{v}$ and $C_{s}$ (proportional to the coupling constants). We studied the system in two scenarios.

In the scenario 1 the parameters $C_s$ and $C_v$ are fixed at $C_s^2=330$ and $C_v^2=249$ for any $\alpha$ and therefore the saturation point--the couple of variables $(\cE_b,\rho_B)_0$ (the binding energy and nucleons density) at which the binding energy, $\cE_b\equiv[\cE(\rho_B)-\rho_BM]/\rho_B$, attains its minimum (ground state) value--varies with $\alpha$. The saturation point for $\alpha=1$ correspond to the physical parameters for nuclear matter, $(\cE_b,\rho_B)_0=(\cE_{b0},\rho_{B0})$. In this scenario the thermodynamic quantities have a strong dependence on $\alpha$ at low temperatures, but in general the system in unstable (Fig. \ref{E_B_T0_2in1} a).

In the second scenario the saturation point is always located at $(\cE_{b0},\rho_{B0})$ and therefore $C_s$ and $C_v$ depend on $\alpha$. In this framework we calculated the relevant physical quantities, such as the binding energy, the effective mass of the nucleons, and the pressure as functions of the state variables, $T$ and $\rho_{B}$. We observe that the thermodynamic quantities are very sensitive to the change of statistics at high temperatures and densities and the system is stable for any $\alpha$.

\section*{Acknowledgments}
This work was supported by the Romanian National Authority for Scientiﬁc Research projects CNCS-UEFISCDI PN-II-ID-PCE-2011-3-0960 and PN09370102/2009. The travel support from the Romania--JINR-Dubna scientific collaboration project, N~4063, is also gratefully acknowledged. 


\end{document}